\documentstyle[12pt,citesort]{article}
\def \el {{\ell}}

\def \K {{\rm K}}

\newcounter{subequation}[equation]

\makeatletter

\def\thesubequation{\theequation\@alph\c@subequation}
\def\@subeqnnum{{\rm (\thesubequation)}}
\def\slabel#1{\@bsphack\if@filesw {\let\thepage\relax
   \xdef\@gtempa{\write\@auxout{\string
      \newlabel{#1}{{\thesubequation}{\thepage}}}}}\@gtempa
   \if@nobreak \ifvmode\nobreak\fi\fi\fi\@esphack}
\def\subeqnarray{\stepcounter{equation}
\let\@currentlabel=\theequation\global\c@subequation\@ne
\global\@eqnswtrue \global\@eqcnt\z@\tabskip\@centering\let\\=\@subeqncr

$$\halign to \displaywidth\bgroup\@eqnsel\hskip\@centering
  $\displaystyle\tabskip\z@{##}$&\global\@eqcnt\@ne
  \hskip 2\arraycolsep \hfil${##}$\hfil
  &\global\@eqcnt\tw@ \hskip 2\arraycolsep
  $\displaystyle\tabskip\z@{##}$\hfil
   \tabskip\@centering&\llap{##}\tabskip\z@\cr}
\def\endsubeqnarray{\@@subeqncr\egroup
                     $$\global\@ignoretrue}
\def\@subeqncr{{\ifnum0=`}\fi\@ifstar{\global\@eqpen\@M
    \@ysubeqncr}{\global\@eqpen\interdisplaylinepenalty \@ysubeqncr}}
\def\@ysubeqncr{\@ifnextchar [{\@xsubeqncr}{\@xsubeqncr[\z@]}}
\def\@xsubeqncr[#1]{\ifnum0=`{\fi}\@@subeqncr
   \noalign{\penalty\@eqpen\vskip\jot\vskip #1\relax}}
\def\@@subeqncr{\let\@tempa\relax
    \ifcase\@eqcnt \def\@tempa{& & &}\or \def\@tempa{& &}
      \else \def\@tempa{&}\fi
     \@tempa \if@eqnsw\@subeqnnum\refstepcounter{subequation}\fi
     \global\@eqnswtrue\global\@eqcnt\z@\cr}
\let\@ssubeqncr=\@subeqncr
\@namedef{subeqnarray*}{\def\@subeqncr{\nonumber\@ssubeqncr}\subeqnarray}
\@namedef{endsubeqnarray*}{\global\advance\c@equation\m@ne%
                           \nonumber\endsubeqnarray}

\makeatletter \@addtoreset{equation}{section} \makeatother
\renewcommand{\theequation}{\thesection.\arabic{equation}}

\def \ci {\cite}

\def \la {\label}

\catcode`\@=11

\newcount\hour
\newcount\minute
\newtoks\amorpm \hour=\time\divide\hour by 60\minute
=\time{\multiply\hour by 60 \global\advance\minute by-\hour}
\edef\standardtime{{\ifnum\hour<12 \global\amorpm={am}%
        \else\global\amorpm={pm}\advance\hour by-12 \fi
        \ifnum\hour=0 \hour=12 \fi
        \number\hour:\ifnum\minute<10
        0\fi\number\minute\the\amorpm}}
\edef\militarytime{\number\hour:\ifnum\minute<10 0\fi\number\minute}

\def\draftlabel#1{{\@bsphack\if@filesw {\let\thepage\relax
   \xdef\@gtempa{\write\@auxout{\string
      \newlabel{#1}{{\@currentlabel}{\thepage}}}}}\@gtempa
   \if@nobreak \ifvmode\nobreak\fi\fi\fi\@esphack}
        \gdef\@eqnlabel{#1}}
\def\@eqnlabel{}
\def\@vacuum{}
\def\marginnote#1{}
\def\draftmarginnote#1{\marginpar{\raggedright\scriptsize\tt#1}}
\def\draft{
        \pagestyle{plain}
        \overfullrule=2pt
        \oddsidemargin -.5truein
        \def\@oddhead{\sl \phantom{\today\quad\militarytime} \hfil
        \smash{\Large\sl DRAFT} \hfil \today\quad\militarytime}
        \let\@evenhead\@oddhead
        \let\label=\draftlabel
        \let\marginnote=\draftmarginnote
        \def\ps@empty{\let\@mkboth\@gobbletwo
        \def\@oddfoot{\hfil \smash{\Large\sl DRAFT} \hfil}
        \let\@evenfoot\@oddhead}

\def\@eqnnum{(\theequation)\rlap{\kern\marginparsep\tt\@eqnlabel}%
        \global\let\@eqnlabel\@vacuum}  }

\renewcommand{\theequation}{\thesection.\arabic{equation}}
\renewcommand{\thefootnote}{\fnsymbol{footnote}}

\def\appendix#1{
  \addtocounter{section}{-2}
  \setcounter{equation}{0}
  \renewcommand{\thesection}{\Alph{section}}
  \section*{Appendix \thesection\protect\indent \parbox[t]{11.15cm}
  {#1} }
  \addcontentsline{toc}{section}{Appendix \thesection\ \ \ #1}
  }
\def \ov {\over}

\def \P { \Phi}

\def \del { \partial}

\def \p {\phi}
\def \ee {\epsilon}
\def \te {\tilde \epsilon}

\def\pd{\partial}

\def\m{\mu}
\def\n{\nu}

\def\g{\gamma}

\def\r{\rho}

\def\s{\sigma}
\def\na{\nabla}

\def\te{\theta} 
\def\ta {\tau}
\def\tta{\tilde{\tau}}
\def\na{\bigtriangledown}
\def\p{\phi}

\def \foot{ \footnote}
\def\be{\begin{equation}}
\def\ee{\end{equation}}
\def \P {\Phi}

\def \ci {\cite}
\def \g {\gamma}

\def \k {\kappa}

\def \m {\mu}
\def \n {\nu}

\def\te{\theta}
\def\g{\gamma}

\begin{document}

\date{}\begin{titlepage}

\begin{center}
\hfill hep-th/0107213\\
\hfill MCTP-01-33\\

\vskip 2.5 cm

{\Large \bf On a Deformation of 3-Branes}

\vskip 1 cm

{\large Leopoldo A. Pando Zayas}

\end{center}

\vskip .6cm
\centerline{\it  Michigan Center for Theoretical Physics}
\centerline{\it Randall Laboratory, Department of Physics}
\centerline{\it The University of Michigan,  Ann Arbor, MI 48109-1120}
\centerline{\tt lpandoz@umich.edu}

\vskip 1.5 cm

\begin{abstract}
We construct an explicit class of solutions of type IIB supergravity
that is a smooth deformation of the 3-brane class of solutions. The
solution is nonsupersymmetric and involves nontrivial dilaton and
axion fields as well as the standard 5-form field strength. One of the
main features of the solution is that for large values of the radius
the deformation is small and it asymptotically approaches the
undeformed 3-brane solution, signaling a restoration of conformal
invariance in the UV for the dual gauge theory. We suggest that the
supergravity deformation corresponds to a massive deformation on the
dual gauge theory and consequently the deformed theory has the
undeformed one as an ultraviolet fixed point. In cases where the
original 3-brane solution preserves some amount of supersymmetry we
suggest that the gauge theory interpretation is that of soft
supersymmetry breaking. We discuss the deformation for D3-branes on
the conifold and the generalized conifold explicitly. We show that the
semiclassical behavior of the Wilson loop suggests that the
corresponding gauge theory duals are confining.
\end{abstract}

\end{titlepage}

\vskip 2.5 cm \vskip 1 cm

\setcounter{page}{1}
\renewcommand{\thefootnote}{\arabic{footnote}}
\setcounter{footnote}{0}

\def \N{{\cal N}}
\def \ov {\over}
\section{Introduction}
One of the main virtues of the AdS/CFT correspondence \cite{ads} is
that it has provided a computationally precise prescription of a
gauge/gravity duality. The original statement of the AdS/CFT refers to
the duality between IIB supergravity on $AdS_5 \times S^5$ and ${\cal
N}=4$ Super Yang-Mills. A very natural question that arises from
phenomenological motivations is the consideration of the gauge/gravity
duality in less supersymmetric situations. With QCD as the ultimate
motivation, one would naturally be interested in nonsupersymmetric,
nonconformal gauge theories. The duals of such gauge theories in the
gauge/gravity framework are expected to be nonsupersymmetric solutions
that do not contain an $AdS_5$ factor in the near horizon limit.

With this main motivation in mind, in this paper we construct a class
of solution of IIB supergravity that is nonsupersymmetric, has running
dilaton and axion, and generically does not contain an $AdS_5$
factor. Our departure from the original 3-brane solution is smooth and
controlled by one parameter. The deformation we construct joins some
examples in the literature where a controlled departure from a
supersymmetric solutions has been motivated from gauge/gravity duality
considerations. In particular, a desire to resolve the singularities
of the KT solution \cite{kt} led to the construction of nonextremal
solutions \cite{nonextremal}. Our class of solutions is closer in
spirit to \cite{sfetsos,gubser,odintsov}, and even more so to
\cite{sfetsos2}, where nonsupersymmetric generalizations have been
found that have an ultraviolet fixed point corresponding to $\N=4$
supersymmetric Yang-Mills.

The construction is based on the well known fact that the super
3-brane\footnote{And for that purpose, any p-brane.} of
\cite{duff,hostro} admits a generalization where the space
perpendicular to the brane is replaced by a Ricci-flat noncompact
space and the warp factor is then a harmonic function on the
perpendicular space. Along these lines, the class of solution
presented in this paper generalizes the situation for the case of
radial-dependent nontrivial dilaton and axion, the warp factor is no
longer a harmonic function in the perpendicular space to the brane but
rather satisfies the Laplace equation with a source.

After discussing the general case in section \ref{statement}, we
consider the case of D3-branes on the conifold. Since the gauge theory
dual of this situation and its generalizations is well understood
\cite{kw,klebanov,ks}, we use this case as the main explicit
example. We suggest that the deformation amounts to adding a
nonsupersymmetric mass deformation to the KW theory. For the
one-parametric deformation we construct, the soft supersymmetry
breaking scale and the confining scale are given by the same
parameter. We therefore, also consider the analogous deformation for
D3-branes on the generalized conifold \cite{pt2} which has a mass
scale in the undeformed theory and is expected to be dual to a
confining gauge theory.

As is common to most of the deformations of p-branes discussed in the
gauge/gravity framework, the resulting deformation we construct has
naked singularities in the IR. The ultimate resolution of this naked
singularity within the full string theory remains an open problem.

\section{Deformed 3-branes}
\label{statement}

The field equations of IIB supergravity when only the metric,
self-dual 5-form, dilaton and axion field are nontrivial are:

\begin{eqnarray}
d * d \P &=& 
e^{2\P} d\chi \wedge {*}d\chi , \nonumber \\
d( e^{2\P} {*}d \chi) &=&0 , \nonumber\\
d{*}\tilde F_5 &=& d F_5 =0,\nonumber\\
R_{MN} &=& {1\over 2} \partial_M \Phi \partial_N \Phi +
{e^{2\P}\over 2} \partial_M \chi \partial_N \chi + {1\over 96}
 F_{MPQRS}  F_N{}^{PQRS}.
\end{eqnarray}
First we consider the dilaton/axion system and show that it
effectively decouples from the specific features of the metric, which
we take to be of the form 
\be
\label{metric}
ds^2=e^{2A(y^m)}\eta_{\m\n}dx^\m dx^\n+e^{2B(y^m)}g_{mn}dy^mdy^n, 
\ee
where $g_{mn}$ is the Ricci-flat metric of the noncompact 6-d space
perpendicular to the 4-d worldvolume. We assume that the space has one
radial coordinate $r$. The typical example to bear in mind would be a
cone over a 5-d compact Einstein space: $dr^2 +r^2 dX_5^2$. However,
the statement is true for more general spaces that are not necessarily
cones over $X_5$; all we need is to be able to extract a volume form
for the 5-d compact space. Assume that the dilaton and axion depend
only on the radial coordinate in the 6-d perpendicular space:
$\P=\P(r)$ and $\chi=\chi(r)$. Then $d\P =\P' dr$ and $d\chi =\chi'dr$
and also 
\be
*d\chi=-\chi' e^{4A+4B}g^{rr}dx^0\wedge dx^1 \wedge dx^2
\wedge dx^3 \wedge Vol(X^5), 
\ee 
where $Vol(X^5)$ is the volume form
of the compact subspace, and we assume it to be independent of the
radial coordinate. Introducing a new radial coordinate $t$ satisfying
\be 
{dt \over dr} =(\sqrt{g} g^{rr})^{-1}e^{-4A-4B}, 
\ee 
the dilaton
and axion equations become:
 
\begin{eqnarray}
\ddot \P&=& \dot \chi^2 e^{2\P}, \nonumber\\
\dot \chi e^{2\P}&=& c_0,
\end{eqnarray}
where $c_0$ is a constant. The dilaton equation becomes $\ddot \P =
c_0^2 e^{-2\P}$, the solution of the dilaton/axion system is:

\begin{eqnarray}
\label{system}
\P&=& -\ln {c_1\over c_0} + \ln \cosh c_1 t, \nonumber \\
\chi&=& \chi_0 + {c_1\over c_0} \tanh c_1 t.
\end{eqnarray}
As was stated above, the decoupling of the dilaton/axion system is
very generic, its functional form is therefore determined
universally. The dependence on the concrete structure of the metric
enters only through $t$ (see \cite{epz} for another example of this
universality).

There are two facts needed in considering the Einstein equations. The
first one states that for a metric of the form (\ref{metric}) the Ricci
tensor has the following form

\begin{eqnarray}
R_{\m\n}&=&-\eta_{\m\n}e^{2(A-B)}\bigg(\na^2 A+ 4 \pd A \pd A + 4 \pd A  \pd B\bigg), \nonumber \\
R_{mn}&=&R_{mn}[g] - g_{mn}\bigg(\na^2 B + 4 \pd B  \pd B +4 \pd A  \pd B\bigg) \nonumber \\
&-& 4 \na_m\na_n B -4 \na_m\na_n A -4 \pd_m A \pd_n A +4\pd_m B \pd_n B \nonumber \\
&+& 4\big(\pd_m A \pd_n B + \pd_n A \pd_m B\big),
\end{eqnarray}
where $R_{mn}[g]$ is the Ricci tensor in the 6-d space calculated from
the metric $g_{mn}$. Motivated by the super 3-brane picture, from now
on we specialize to the case $A+B=0$; in the absence of the
deformation this condition is required for supersymmetry
\cite{duff,kehagias}. Moreover, this choice is consistent with the
equations of motion. The ansatz for the 5-form field strength is

\be
F_5= {\cal F}+ *{\cal F}, \, \, \,
 {\cal F}=\K Vol(X^5).
\ee
The $F_5$ field equation (which coincides  with its Bianchi identity)
requires $\K=\rm{Const.}$. Note that this is different from the
standard 3-brane solution, where $F_5$ can also be, and usually it is,
written as $F_5 = (1+*)dh^{-1}\wedge dx^0 \wedge dx^1 \wedge dx^2
\wedge dx^3$, where $h=e^{-4A}$. Actually, in our case

\be 
*{\cal F}={\K \over h^2 g^{rr}\sqrt{g}}dr\wedge dx^0 \wedge dx^1
\wedge dx^2 \wedge dx^3.  
\ee 
The last element needed for the Einstein
equation is the observation that the energy momentum tensor of the
dilaton/axion system contributes a ``charge''. Namely, using the
solution of the dilaton/axion system given in equation (\ref{system})
one obtains

\be 
{1\over 2}\P'^2 +{1\over 2} e^{2\P}\chi'^2 = {1\over 2}c_1^2
({dt\over dr})^2 .  
\ee 
There are two independent Einstein equations
(one of them is the $rr$ component)

\begin{eqnarray}
g_{rr}\na^2A-8A'^2&=&-{\K^2 e^{8A}\over 4g(g^{rr})^2}+ {1\over 2}{c_1^2 \over g (g^{rr})^2}, \nonumber \\
\na^2 A&=&{\K^2e^{8A}\over 4 g g^{rr}}.
\end{eqnarray}
Combining these two, we obtain an equation for $A$
\be 
\na^2A-4g^{rr}A'^2 = {c_1^2\over 4 g g^{rr}}.
\ee
Written in terms of the coordinate $t$ and the warp factor $h$, this
is simply the equation of the harmonic oscillator

\be 
\ddot h + c_1^2 h=0.  
\ee Note that for trivial dilaton/axion
system we recover the typical 3-brane solution $h=h_0 + h_1 t$. The
other independent equation fixes the value of $\K$.  Thus, we have
shown that the type IIB field equations admit a generalization of the
3-brane solution of the form:

\begin{eqnarray}
ds_{10}^2 &=& h^{-1/2}(r)\eta_{\m\n} dx^\m dx^\n + h^{1/2}(r)g_{mn}dy^m dy^n, \nonumber\\
h&=&b_1 \sin c_1 t + b_2 \cos c_1 t, \nonumber \\
F_5&=& (1+*)\K\,\, Vol(X^5), \nonumber \\
\P&=& -\ln {c_1\over c_0} + \ln \cosh c_1 t, \nonumber \\
\chi&=& \chi_0 + {c_1\over c_0} \tanh c_1 t,
\end{eqnarray}
where $t$ and $r$ are related through  $dt/dr =(\sqrt{g}g^{rr})^{-1}$. 

The limit necessary to recover the standard 3-brane solution involves
sending $c_1 \to 0$ and $ b_1 \to\infty$ in such a way that $c_1 b_1
\to \rm{const.}$ Note also that in a region where $t\to 0$ the
solution tends to the standard 3-brane solution.  The solution is not
supersymmetric. This is easily seen from the supersymmetric variation
of the dilatino:

\be
\delta \lambda = -{1\over 2\tau_2}\g^M\pd_M\tau \, \epsilon^{*}= -{1\over2} e^{\P} \g^r (\pd_r \chi + i \pd_r e^{-\P}) \epsilon^{*},
\ee
which can not be made zero \footnote{There is, however, a different
situation in the Euclidean version as discussed in \cite{sfetsos2}.}
for nontrivial values of the $\P$ and $\chi$.

\section{Deformed D3-branes on Conifolds}
In this section we consider the above construction in the particular
case where the 6-d perpendicular space is the conifold. This example
will clarify any technical detail left obscure in the general
construction of section \ref{statement} and, more importantly, will
shed light on the physics of the deformed solution. This a
phenomenologically very interesting case and it also captures most of
the general properties of the solution\footnote{The study of this
example motivated the general construction presented in section
\ref{statement}. I thank A.A. Tseytlin for originally suggesting this
example for consideration.}. The metric of the singular conifold is

\be
\label{conifold}
ds_6^2=dr^2 + r^2\bigg[{1\over 9} e_\psi^2 + {1\over 6} \big( e_{\te_1}^2 +
e_{\phi_1}^2 + e_{\te_2}^2 + e_{\phi_2}^2\big)\bigg],
\ee
where
\be e_{\theta_i}=d\theta_i, \, \quad  e_{\phi_i}=
\sin\theta_id\phi_i,\, \quad
 e_{\psi} = d\psi +  \cos \te_1 d\p_1  +  \cos \te_2 d\p_2  \ .
\ee
All the geometrical statements made in this section apply as well to
any 6-d space that is a cone over a 5-d compact Einstein manifold.
The explicit form for the selfdual 5-form is
\be
F_5= {\cal F}+ *{\cal F}, \, \, \,
 {\cal F}=\K(\r) e_{\psi}\wedge
e_{\theta_1}\wedge e_{\phi_1}\wedge e_{\theta_2}\wedge e_{\phi_2}\ .
\ee
At the expense of being repetitive but with the hope of completely
clarifying the structure of the Einstein equation we consider them
explicitly in this case. The two independent Einstein equations are

\begin{eqnarray}
A'' + {5A'\over r}& = &{108 K^2 e^{8A}\over r^{10}}, \nonumber \\
A'^2 &=&{ 27K^2e^{8A}\over r^{10}}-{1\over 16}(\P'^2 + e^{2\P} \chi'^2).
\end{eqnarray}
The equation for $A$ becomes
\be
A''+{5A'\over r}-4A'^2 = {c_1^2\over 4 r^{10}}.
\ee
To solve this equation take $h=e^{-4A}$ and note that
$$
(r^5 (e^{-4A})')'=-4r^5e^{-4A}(A''+{5A'\over r}-4A'^2),
$$
which linearizes it into
\be
h''+{5h'\over r} + {c_1^2 h\over r^{10}}=0.
\ee
In the $t$ coordinate $t$: $\,\,\, dt/dr = r^{-5}$ this equation becomes
\be
\ddot h + c_1^2 h =0.
\ee
Note that in the absence of the deformation,  $\ddot h=0$ implies that  $h$ has the form of a 
a harmonic function:       $h= h_0 + h_1 t= h_0 + Q_1/r^4$.
The solution of the whole system, setting  $c_1 = 4L^4$, is

\begin{eqnarray}
e^\P&=& g_s  \cosh {L^4\over r^4}, \nonumber \\
\chi&=& -g_s^{-1}\tanh {L^4\over r^4}, \nonumber \\
h&=& b_1 \sin{L^4\over r^4}+  b_2\cos {L^4\over r^4}, \nonumber \\
K &=& {L^4\over 27} \sqrt{b_1^2 + b_2^2}.
\label{defd3con}
\end{eqnarray}

Taking $b_2=1$ in the above expressions gives asymptotically flat
space far away from the brane. The other interesting asymptotic to
explore corresponds to removing the flat region, i.e. taking $b_2 =
0$.  The limit of D3 without deformation can be recovered by sending
$L^4 \to 0$ and $ b_1 \to\infty$ in such a way that $L^4 b_1 \to
\rm{const.}$ The trigonometric behavior of the metric signals the
existence of a very pathological array of naked singularities. We
thus, consider the solution reliable away from this region. Namely, to
avoid the highly oscillatory $r \sim 0$ region we assume $r\ge r_*$,
where $r_*$ is the largest value of $r$ at which $h$ has a
maximum. This pathological behavior is captured by the scalar
curvature which equals

\be
R={8L^8\over h^{1/2}r^{10}}.
\ee
Showing that zeros of $h$ are points at which $R$ blows up. In the
limit of no deformation we have that, as expected for 3-branes, the
scalar curvature is identically zero. This behavior of $R$ is then
solely an effect of the dilaton/axion system being nontrivial.

Note that now, in contrast to the standard D3-brane, the 10-d metric
does not factorize into a product metric after removing the
asymptotically flat region. Removing the asymptotically flat region
(taking $b_2=0$) for the deformed 3-brane leaves the $\sin(L^4/r^4)$
factor in the angular part, preventing the metric from becoming a
product space.  One, however, recovers the $AdS_5 \times T^{1,1}$
space in the limit of no deformation discussed above. Also, very far
from the brane $r\gg L$ the deformation becomes asymptotically
small. Thus, in terms of the AdS asymptotics, our solution tends to
$AdS_5\times T^{1,1}$ for large values of the radius (which
corresponds to the UV region on the gauge theory side). As one
approaches small r (flow to the IR in gauge theory) the metric is no
longer a direct product. Moreover, in order to avoid a very erratic
behavior (the warp factor going to zero and even changing signs) we
should consider our solution being valid only in the region $r \ge
r_*$ where $r_*$ is the largest value of $r$ at which $h$ has a
maximum. We suggest that around $r\approx r_*$ the supergravity
description starts to break down. The asymptotic behavior is very
different from the KT solution in the sense that KT is not
asymptotically $AdS_5 \times T^{1,1}$, there is the characteristic
logarithmic running of the warp factor \cite{kt} which is also present
for all of its generalizations \cite{ks,pt1,pt2}. Perhaps the most
important difference with the KT solution is the fact that for our
solution $\K$ is constant, i.e. no running 5-form flux (or number of
colors).  This allows us to conclude that the deformation does not
alter the gauge group structure of the KW solution. In other words,
the deformation amounts to a perturbation that dies in the UV but has
a very strong influence in the IR. The situation described above is
the trademark of a massive deformation. This situation is somehow
analogous to that considered for holomorphic $\tau=\chi + ie^{-\P}$
\cite{mello,pg}. One difference is, however, supersymmetry. The
solution presented above is not supersymmetric, as was shown above
from the supersymmetric transformation of the dilatino. In the absence
of nontrivial 3-form field strengths the complex dilaton $\tau$ has to
be a holomorphic or antiholomorphic function (see for example
\cite{epz}).Thus, our deformation amounts, in the gauge theory dual,
to a nonsupersymmetric mass deformation.  Since supersymmetry is
restored at high energy we can conclude that this is a situation where
supersymmetry is softly broken. Recalling that the ${\cal N}=1$
superconformal field theory (the limit with the mass deformation going
to zero) is known \cite{kw,klebanov,ks}, we suggest that the KW gauge
theory is a UV fixed point of the deformed theory; and also that the
deformation is given by adding dimension two operators of the form
$m^2\rm{Tr}( a^{\dagger} a + b^{\dagger} b + a b)$, where $a$ and $b$
are the components of the chiral superfields $A$ and $B$ transforming
as $(N, \bar N) $ and $(\bar N, N)$ respectively under the gauge group
$U(N) \times U(N)$. For the gauge couplings one has:

\be
{1\over g_1^2}+ {1\over g_2^2}  \sim  e^{-\P}= {1\over g_s  \cosh {L^4\over r^4}}.
\ee
The difference of the squares of the inverses of the gauge couplings
is proportional to $B_2$ which is constant for this solution.  The
behavior of the gauge couplings is moderate, the above sum ranges from
$1/ g_s$ to $1/(g_s \cosh L^4/r_*^4)$.

The dependence of the dilaton on the radius is very mild in the region
 where the solution is reliable, even after applying a $SL(2,R)$
 transformation to the above background. Namely:

\be
\tta = {a\ta + b \over c\ta +d}, \quad \ta=\chi + i e^{-\P}, \quad
\left(
\begin{array}{cc}
a&b\\
c&d\\
\end{array}
\right)\in SL(2,R),
\ee
 results in the following dilaton
\be
e^{\tilde \P}=g_s\bigg[{c^2\over g_s^2}+ d^2 -{2cd\over g_s}\tanh{L^4\over r^4}\bigg] \cosh{L^4\over r^4}.
\ee
It is worth noting that for $r\gg L$ we basically recover the
strong/weak relation. If the original string coupling $g_s$ was weak
($g_s\ll 1)$, then after the $SL(2,R)$ transformation we get $\tilde
g_s \approx c^2/g_s \gg 1$.


\subsection{Deformed D3-branes on the  generalized  conifold }

In the previous subsection we considered a deformation of the
background of D3-branes on the singular conifold. One feature of the
solution was that there was only one scale which determined, on the
gauge theory side, the confining (see appendix) and the supersymmetry
breaking scale.  In this subsection we consider adding a deformation
to the background of D3 on the generalized conifold \cite{pt2}.  The
generalized conifold is not a cone over a 5-d Einstein
space. Nevertheless, the general discussion of section \ref{statement}
directly applies since we are able to define a volume form for the
compact 5-d space that is independent of the radial coordinate,
basically the same one as in the case of the singular
conifold\footnote{Similarly, one can define such volume form for the
resolved and deformed conifolds.}. D3-branes on the generalized
conifold provide supersymmetric vacua of IIB \cite{papats}, and they
are dual to a confining theory \cite{pt2}. Thus, D3-branes on the
generalized conifold provide a supersymmetic massive deformation of
KW. By considering the axion/dilaton deformation on the background of
D3 on the generalized conifold we believe, in principle, to be able to
separate the confining and susy breaking scales in the gauge theory by
adjusting the two corresponding parameters in the sugra solution.

The metric of the generalized conifold is:
\be
ds^2 = \k^{-1} 	dr^2 + \k {r^2\over 9} e_\psi^2 + {r^2\over 6} \big( e_{\te_1}^2 +
e_{\phi_1}^2+ e_{\te_2}^2 + e_{\phi_2}^2\big),
\ee
with 
\be
\k=1-{b^6\over r^6}.
\ee
The  radial coordinate $t$ is given by 
\be
{dt\over dr}= {1\over \kappa r^5}.
\ee
The  solution is 
\be
e^\P=g_s \cosh c_1 t, \qquad \chi = g_s^{-1}\tanh c_1 t .
\ee
and 
\be
h=b_1 \sin c_1 t + b_2 \cos c_1 t, \quad  \rm {and} \quad K={c_1\over 108} \sqrt{b_1^2
+b_2^2}.
\ee
The relation between the radial coordinate $t$ and the original $r$ is
\be
t={1\over 2b^4}\bigg[{1\over 6}\ln {(\bar r^2-1)^3\over \bar r^6-1} + {1\over
\sqrt{3}}({\pi\over 2}-\arctan{2\bar r^2 +1\over \sqrt{3}})\bigg],
\ee
where $\bar r = r/b$. In the  $r\gg b$ limit we basically recover the previous
singular conifold situation $t\to -1/(4 r^4)$. The generalized conifold metric requires
$r\ge b$. In the $r\to b$ limit we have
$$
t\to {1\over 6b^4}\ln ({r\over b} -1).
$$
This behavior means that the restriction $r\ge b$ is not good enough
since by the time we reach $r\approx b$ the metric has experienced
some naked singularities $h=0$. We need, therefore, to restrict the
radial domain further. As in the case of the conifold, we take $r\ge
r_*$ where $r_*$ is the largest values of $r$ for which $h$ has a
maximum.

\section*{Acknowledgments}
I am especially thankful to A.A. Tseytlin for originally suggesting
the deformation of the conifold for consideration, as well as many
other comments and suggestions that considerably improved this
paper. I also benefited from discusions with D. Chung,
M. Einhorn and J. Liu. I would like to acknowledge the Office of the
Provost at the University of Michigan and the High Energy Physics
Division of the Department of Energy for support.

\appendix{Wilson Loop behavior}

Let us now investigate, following \cite{maldacena,rey}, the behavior
of the Wilson loop corresponding to the ``quark-antiquark" potential
in the dual gauge theory. It is given by the exponential of the
classical fundamental string action in this background evaluated for a
static configuration of open string ending on the probe D3-brane
placed at the ``boundary" $r= \infty$.

We will show that generically it is possible to obtain an area law
(confining) behavior for the deformed D3-brane background.  The
existence of an area law reinforces the interpretation of the
deformation as a massive deformation, since otherwise dimensional
analysis will require the potential to be Coulombic\footnote{I thank
M. Einhorn for useful discussions on this appendix.}. This is
different from what is found in the standard conifold case \ci{kw}
where the potential is Coulombic as in \cite{maldacena,rey} in any
regime. We will also show that, as expected, in the limit when the
deformation is turned off we recover the Coulombic potential
characteristic of D3-branes.

The relevant metric is the string metric rather than the Einstein
metric of equation (\ref{defd3con}), $ ds_{string}^2={e^{\P/2}\over
g_s^{1/2}}ds_{Einstein}^2$, which in our background is

\be
ds_{st}^2=\left(\cosh{L^4\over r^4}\right)^{1/2}\bigg[h^{-1/2}\eta_{\m\n}dx^\m dx^\n +h^{1/2}ds_6^2\bigg],
\ee
where $ds_6^2$ is the conifold metric (\ref{conifold}) and $h=b_1\sin
L^4/r^4$ in order to remove the asymptotically flat region. The
Nambu-Goto string action which determines the expression for the
Wilson loop depends on this 10-d string metric $G_{MN}$ as $\int d\ta
d\s \sqrt{-\det (G_{MN}\del_a X^M\del_b X^N)}$.  In the static gauge
($x_0= \tau, \ x_1\equiv x= \s$) and assuming that the string is
stretched only in the radial direction, i.e. only the $r$ coordinate
depends on $\s$, we get\foot{$T$ is the time interval and the string
tension is set equal to 1.} \be S=T\int dx
\sqrt{G_{00}G_{xx}+G_{00}G_{rr}(\pd_x r)^2} = T\int dx e^{\P/2}
\sqrt{h^{-1}+(\pd_x r)^2}\ .  \ee Since the Lagrangian of this
``mechanical system" does not depend explicitly on the ``time" $x$, we
have a conserved quantity $ {e^{\P/2}h^{-1}\over \sqrt{h^{-1}+(\pd_x
r)^2}}=e_0 $. The natural way to parametrize $e_0$ is by interpreting
it as the total energy of the ``mechanical system'' and therefore
relating it to the turning point $r_0$ . Then \be
e_0^2={\cosh{L^4\over r_0^2} \over b_1 \sin {L^4\over r_0^4}}.  \ee
From the above expressions we find

\be 
dx= {h^{1/2}dr\over \sqrt{e^\P/( e_0^2 h)-1}}= 
{b_1^{1/2}\sin^{1/2}(L^4/r^4)\over \sqrt{(\cosh (L^4/r^4)/( e_0^2 b_1\sin (L^4/r^4))-1}}dr  \ .
\la{xxx} 
\ee
 The energy
of a  static string configuration is thus 
\be 
E={S\over T}= \int dx e^{\P/2}
\sqrt{h^{-1}+(\pd_x r)^2}= \int {e^{\P/2} \over e_0 h^{1/2}}{dr \over \sqrt{e^\P/( e_0^2 h) -1}} . 
\ee 

Following \cite{maldacena,rey}, the question about confinement is then
 reduced to finding the dependence of the energy $E$ on the distance
 $\el$ between the string end-points (between the ``quark" and the
 ``antiquark"). More precisely, it has been discussed extensively (see
 for example \cite{petrini}), that the behavior of the Wilson Loop is
 essentially determined by the properties around the turning point$r_0$ of
 the probe string. Following that prescription we obtain

\be
\el \approx {2b_1^{1/2}\sin^{1/2} (L^4/r_0^4)\, r_0^3 \over L^2\sqrt{\cot(L^4/r_0^4)-\tanh (L^4/r_0^4)}}, \quad 
E\approx {r_0^3 \over L^2 \sqrt{\cot(L^4/r_0^4)-\tanh(L^4/r_0^4)}}.
\ee

It is worth noting that in the limit of zero deformation we reproduce
the Coulombic behavior of \cite{maldacena,rey}.  Namely, taking
$L^4\to 0, \,\, b_1 \to \infty$ with $b_1L^4=R^4$ we obtain

\be
\el \approx {2R^2\over r_0}, \qquad E\approx r_0,
\ee
which is precisely the Coulombic behavior $E \approx
2R^2/\el$. Recalling that taking the zero deformation limit and
considering the region far away from the brane are formally the same
we see that if $r_0$ is very large (that is the string does not probe
deep into the bulk) we also obtain a Coulombic potential. More
generally, however, we have

\be 
E\approx {1\over 2 b_1^{1/2} \sin^{1/2}(L^4/r_0^4)}\, \el.  
\ee 
In the above expression $r_0(\el)$ but we find almost linear confinement
whenever the $\sin$ is not small, which is everywhere except in the
previously discussed limit. In particular in the region $r_0 \sim L
\sim \el$ whe obtain that $E\approx \el$.

\end{document}